\documentclass{elsart}
\usepackage{graphicx}
\usepackage{dcolumn,color}
\usepackage{bm}
\usepackage{epsfig}
\usepackage[english]{babel}

\begin{document}                                                               
\bibliographystyle{try}
\topmargin 0.1cm
\vspace*{10mm}
\begin{center}
{\large\bf THE GLUEBALL CANDIDATE $\bf\eta(1440)$ \\
\vskip 3mm
AS $\bf\eta$ RADIAL EXCITATION
}

\vspace*{10mm}
{\sc Eberhard Klempt}

\vspace*{2mm}
{\it Helmholtz-Institut f\"ur Strahlen-- und Kernphysik \\
	Universit\"at Bonn\\
	Nu{\ss}allee 14-16, D-53115 Bonn, GERMANY\\
	\texttt{e-mail: klempt@hiskp.uni-bonn.de}
        }
\end{center}
\vspace*{20mm}
Abstract\\
The Particle Data Group decided to split the $\eta(1440)$ into
two states, called $\eta_L$ and $\eta_H$. The $\eta(1295)$ 
and the $\eta_H$ are supposed to be the radial excitations of the
$\eta$ and $\eta'$, respectively. The $\eta_L$ state cannot be
accomodated in a quark model; it cannot be a $q\bar q$ state, 
however, it might be a glueball. In this contribution it is shown
that that the $\eta(1295)$ does not have the properties which must
be expected for a radially excited state. The splitting of the
$\eta(1440)$ is traced to a node in the wave function of a 
radial excitation. Hence
the two peaks, $\eta_L$ and $\eta_H$, originate from one resonance
which is interpreted here as first radial excitation of the $\eta$.
\vfill
\begin{center}
Contributed to \\
32nd International Conference on High Energy Physics\\
August 16 -- 22,  2004\\
Beijing, China\\
\end{center}
\vspace*{10mm}
\clearpage

\section{Short history of the $\eta (1440)$}
The E/$\iota$ was discovered in 1967 in $p\bar p$
annihilation at rest into $(K\bar K\pi)\pi^+\pi^-$.
It was the first meson found in a European experiment  
and was called E-meson~\cite{Baillon67}. 
Mass and width
were determined to be $M = 1425 \pm 7, \Gamma = 80\pm 10$\,MeV, 
with quantum numbers $J^{PC} = 0^{-+}$.
In the charge exchange reaction $\pi^- p \to n\rm K\bar K\pi$, 
using a 1.5 to 4.2\,GeV/c pion beam~\cite{Dahl:ad},  
a state was observed with 
$M = 1420 \pm 20, \Gamma = 60\pm 20$\,MeV and
$J^{PC} = 1^{++}$. Even though the quantum
numbers were different, it was still called E-meson.
\par
In 1979 there was a claim~\cite{Stanton:ya} 
for the $\eta (1295)$ which was later
confirmed in other experiments.
In 1980 the E--meson was observed~\cite{Scharre:1980zh} 
in radiative J/$\psi$ decays into
$(K\bar K\pi)$ with $M = 1440 \pm 20, \Gamma = 50\pm 30$\,MeV;
the quantum numbers were `rediscovered'~\cite{Edwards:1982nc} 
to be $J^{PC} = 0^{-+}$. The E--meson was renamed $\iota (1440)$ to
underline the claim that it was the $\iota^{\rm st}$ glueball
discovered in an experiment. 
The $\iota (1440)$ is a very strong signal, one of the strongest, in
radiative J/$\psi$ decays.
The radial excitation $\eta (1295)$
is not seen in this reaction; hence the  $\iota (1440)$
must have a different nature. At that time it was
proposed (and often still is) to be a glueball. 
Further studies, in particular by the Obelix collaboration at 
LEAR~\cite{Nichitiu:2002cj},
showed that the $\iota (1440)$
is split into two components, a
$\eta_L\to a_0(980)\pi$ with $M = 1405 \pm 5, \Gamma = 56\pm 6$\,MeV
and a $\eta_H\to \rm K^*\bar K +\bar K^*K$ with $M = 1475 \pm 5,
\Gamma = 81\pm 11$\,MeV: there seem to be 3 $\eta$ states
in the mass range from 1280 to 1480\, MeV.
\par
The $\eta (1295)$ is then likely the radial excitation of the
$\eta$. It is mass degenerate with the $\pi (1300)$, hence
the pseudoscalar radial excitations seem to be ideally mixed.
Then, the $\bar ss$ partner should have a 
240 MeV higher mass. The $\eta_H$ could play this role.
The $\eta_L$ does not find $\eta_L$ a slot in the spectrum of 
$\bar qq$ mesons;
the low mass part of the $\iota (1440)$ could be a glueball.
This conjecture is consistent with the observed decays.
A pure flavor octet $\eta (xxx)$ state decays into $\rm K^*K$ but
not into $a_0(980)\pi$. In turn, a pure flavor singlet $\eta
(xxx)$ state decays into $a_0(980)\pi$ but not into $\rm K^*K$.
The $\eta_H$, with a large coupling 
to $\rm K^*K$, cannot 
possibly be a glueball, whereas the $\eta_L$ with its
$a_0(980)\pi$ decay mode can be.
\par
The PDG 2004 supports this interpretation
of the pseudoscalar mesons~\cite{Eidelman:2004wy}:
\\
\begin{center}
\hspace*{-6mm}\fbox{\fbox{
\begin{tabular}{ccccc}
$\pi$        & $\eta$        &             & $\eta^{\prime}$ & $K$ \\
$\pi(1300)$  & $\eta(1295)$  & $\eta(1405)$ & $\eta(1475)$ & $K(1460)$ \\
$n\bar n$  & $n\bar n$  & {glueball}     & $s\bar s$ & $n\bar s$  \\ 
\end{tabular}
}}
\end{center}
\vskip 3.5mm
\par
Two quantitative tests have been proposed to
test if a particular meson is glueball--like:
the stickiness and the gluiness.
The stickiness of a resonance R with mass $m_{\rm R}$ and two--photon width $\Gamma _{{\rm R} \to \gamma\gamma}$ 
is defined as:
$$
S_{\rm R} = N_l \left(\frac{m_{\rm R}}{K_{{\rm J}\to\gamma {\rm R}}}\right)^{2l+1}
\frac{\Gamma _{{\rm J}\to\gamma {\rm R}}}{\Gamma _{{\rm R} \to \gamma\gamma}} \ ,
$$
where $K_{{\rm J}\to\gamma {\rm R}}$ is the energy of the photon  in the J rest frame, 
$l$ is the orbital angular momentum of the two initial photons or gluons ($l=1$ for $0^-$), 
$\Gamma _{{\rm J}\to\gamma {\rm R}}$ is the J radiative decay width for R, 
and $N_l$ is a normalization factor chosen to give $S_{\eta} = 1$.
The L3 collaboration determined~\cite{Acciarri:2000ev} this 
parameter to be $S_{\eta(1440)}=79\pm 26$.

The gluiness ($G$) was introduced~\cite{Close:1996yc,Paar:pr}
to quantify the ratio of the two--gluon and two--photon coupling of a particle
and is defined as:
$$ 
G = \frac{9\,e^4_q}{2}\,\biggl(\frac{\alpha}{\alpha _s}\biggr)^2 \,
\frac{\Gamma _{{\rm R} \to {\rm gg}}}{\Gamma _{{\rm R} \to \gamma\gamma}} \ ,
$$
where $e_q$ is the relevant quark charge. $\Gamma _{{\rm R} \to {\rm
gg}}$ is the two--gluon width of the resonance {\rm R}, 
calculated from equation (3.4) of ref.~\cite{Close:1996yc}.
Stickiness is a relative measure, gluiness is a normalised quantity
and is expected to be near unity for a $q\bar{q}$ meson.
The L3 collaboration determined~\cite{Acciarri:2000ev} this quantity,  
$G_{\eta(1440)}=41\pm 14$.
\par
These numbers can be compared to those for the $\eta '$ for which 
$S_{\eta '} = 3.6 \pm 0.3$ and $G_{\eta '} = 5.2 \pm 0.8$ 
is determined, for $\alpha_s(958 MeV)=0.56\pm0.07$. 
Also $\eta'$ is `gluish', but much more the $\eta_L$.
The $\eta_L$ is the first glueball\,!

\section{The $\eta (1295)$ and the $\eta (1440)$ in radiative J/$\psi$
decays}
Radiative J/$\psi$ decays show an asymmetric peak in the $\eta(1440)$
region therefore both the  $\eta_L$ and the $\eta_H$, must contribute to
the process. Obvoiusly, radial excitations are produced in radiative 
J/$\psi$ decays (not only glueballs). The $\eta(1295)$ must therefore
also be produced, but it is not - at least not with the expected
yield. Is there evidence for this state
in other reactions\,?
\par
At BES, $\eta (1295)$ and $\eta (1440)$ were studied in
J/$\psi\to(\rho\gamma)\gamma$ and $\to
(\phi\gamma)\gamma$~\cite{Bai:2004qj}. The
$\eta(1440)$ (seen at 1424\,MeV) is seen to decay strongly 
into $\rho\gamma$ and not into $\phi\gamma$. This is not consistent
with the hypothesis of $\eta(1475)$ being a $s\bar s$ state. A peak 
below 1300\,MeV is assigned to the $f_1(1285)$ even though a small
contribution from $\eta(1295)$ cannot be excluded.  
\section{The $\eta (1295)$ and the $\eta (1440)$ in $\gamma\gamma$ at LEP}
Photons couple to charges; in $\gamma\gamma$ fusion
a radial excitation is hence expected to
be produced more frequently than a
glueball. In $\gamma\gamma$ fusion, both electron and positron scatter
by emitting a photon. If the momentum transfer to
the photons is small, the $e^+$ and $e^-$ are scattered into
forward angles (passing undetected through the beam pipe), thus
the two photons are nearly real. If the $e^+$ or $e^-$
has a large momentum transfer, the photon acquires mass, and we
call the process $\gamma\gamma^*$ collision. Two massless
photons couple to the $\eta(1295)$ but not to the 
$ f_1(1285)$; in this way, a peak at $\sim$1290\,MeV can be
identified as one of the two states. The L3 collaboration studied 
$\rm\gamma\gamma^*$ and $\gamma\gamma\to K^0_sK^{\pm}\pi^{\mp}$.
At low $q^2$, a peak at 1440\,MeV is seen, it requires high
$q^2$ to produce a peak at 1285\,MeV. A pseudoscalar
state is produced also at vanishing $q^2$ while $J^{PC}=1^{++}$
is forbidden for $q^2\to 0$. Hence the structure at 1285\,MeV
is due to $f_1(1285)$ and not due to $\eta(1295)$. There is
no evidence for $\eta (1295)$ from $\gamma\gamma$ fusion.
The stronger peak contains contributions from    
$\eta (1440)$ and  $f_1(1420)$~\cite{Acciarri:2000ev}.
The coupling of
the $\eta$(1440) meson to photons is stronger than that of the $\eta$(1295): 
the assumption that the $\eta$(1295) is a $(u\bar u+d\bar d)$
radial excitation must be wrong\,!

\section{The $\eta (1295)$ and $\eta (1440)$ in $p\bar p$ annihilation}
\begin{figure}[h!]
\begin{center}
\begin{tabular}{cc}
\hspace*{-2mm}\epsfig{file=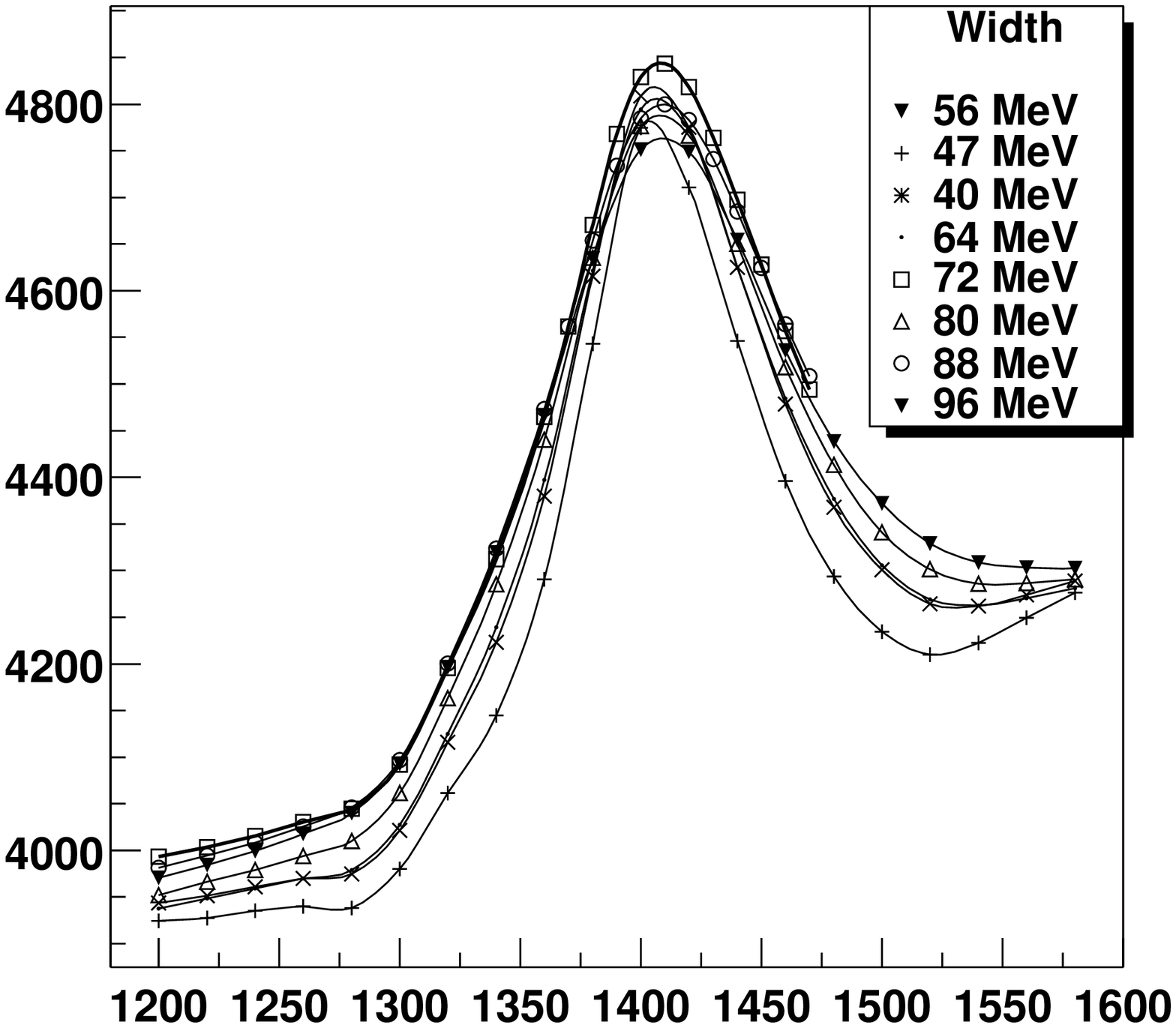,width=0.5\textwidth} &
\hspace*{-4mm}\epsfig{file=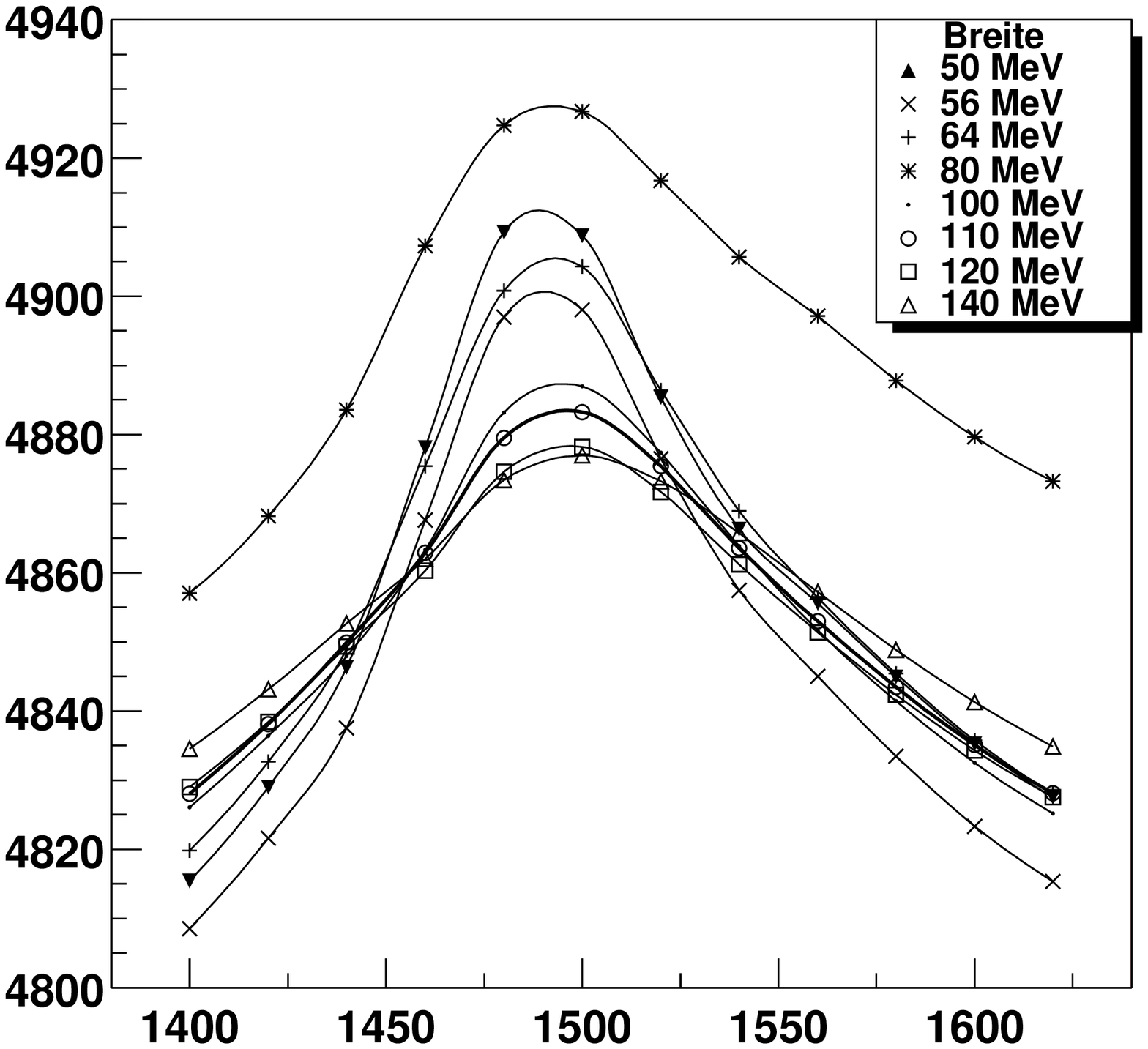,width=0.5\textwidth}
\end{tabular}
\end{center}
\caption{\label{escan}
Scan for a $0^+ 0^{- +}$ resonance with
 different widths\protect~\cite{Reinnarth}. The likelihood optimizes for
$M=1407\pm 5, \Gamma = 57\pm 9$\,MeV. The resonance 
is identified with the $\eta_L$.  
A search for a second pseudoscalar resonance (right panel)
gives evidence for the $\eta_H$ with 
$M=1490\pm 15, \Gamma = 74\pm 10$\,MeV.}
\end{figure}

The Crystal Barrel collaboration searched for the 
$\eta (1295)$ and $\eta (1440)$ in
the reaction $p\bar p\to\pi^+\pi^-\eta (xxx)$,  
$\eta (xxx)\to\eta\pi^+\pi^-$. The search was done by assuming the presence
of a pseudoscalar state of given mass and width, mass and width
are varied and the likelihood of the fit is plotted.
Fig.~\ref{escan} shows such a plot~\cite{Reinnarth}.
A clear pseudoscalar resonance signal is seen 
at 1405\,MeV. Two decay modes are observed, $a_0(980)\pi$ and
$\eta\sigma$ with a ratio $0.6\pm0.1$. We use the notation 
$\sigma$ for the full $\pi\pi$ S--wave. 
\par
A scan for an additional $0^+ 0^{- +}$ resonance 
provides no evidence for the $\eta (1295)$ but for a second
resonance at 1480\,MeV, see Fig.~\ref{escan}, with $M=1490\pm 15
,\Gamma=74\pm 10$. This is the $\eta_H$. It decays to
$a_0(980)\pi$ and $\eta\sigma$ with a ratio $0.16\pm0.10$. 
This data provides the first evidence for $\eta_H\to\eta\pi\pi$ decays. 

The phenomena observed in the pseudoscalar sector are confusing:
The $\eta (1295)$, the assumed radial excitation of the $\eta$, is
only seen in $\pi^- p\to n (\eta\pi\pi)$, not in $p\bar p$
annihilation, nor in radiative J/$\psi$ decay, nor in
$\gamma\gamma$ fusion. In all these reactions it should have been 
observed. There is
no reason for it to have not been produced if it is a $\bar qq$ state. On
the other hand, we do not expect glueballs, hybrids or multiquark
states so low in mass. In the 70's, the properties of the
$a_1(1260)$ were obscured by the so--called Deck effect
($\rho$--$\pi$ re-scattering in the final state). Possibly,
$a_0(980)\pi$ re-scattering fakes a resonant--like behavior but
the $\eta(1295)$ is too narrow to make this possibility realistic.
Of course there is the possibility that the $\eta(1295)$ is
mimicked by feed--through from the $f_1(1285)$. In any case, I
exclude the $\eta (1295)$ from the further discussion.
\par
The next puzzling state is the $\eta (1440)$. It is  not
produced as $\bar ss$ state but decays with a large fraction
into $\rm K\bar K\pi$ and it is split into two components.
I suggest that the origin of these anomalies is due
to a node in the wave function of the $\eta(1440)$\,!
This node has an impact on the decay matrix elements 
calculated by Barnes {\it et al.}~\cite{Barnes:1996ff} 
within the $^3P_0$ model.
\par
\section{$E/\iota$ decays in the $^3P_0$ model}
The matrix elements for decays of the $\eta (1440)$ as a
radial excitation (=$\eta_R$) depend on spins, parities and decay
momenta of the final state mesons. For
$\eta_R$ decays to $\rm K^*K$, the matrix element is given by
$$
f_P = \frac{2^{9/2}\cdot 5}{3^{9/2}}\cdot x\left(1-\frac{2}{15}x^2 \right).
$$
In this expression, $x$ is the decay momentum in units of 400\,MeV/c;
the scale is determined from comparisons of measured partial widths
to model predictions. The matrix element vanishes for $x=0$ and
$x^2 = 15/2$, or $p=1$\,GeV/c. These zeros have little effect on the
shape of the resonance.
\par
\begin{figure}[h!] 
\begin{tabular}{ccc} 
\hspace*{-4mm}\includegraphics[width=0.36\textwidth,height=6cm]{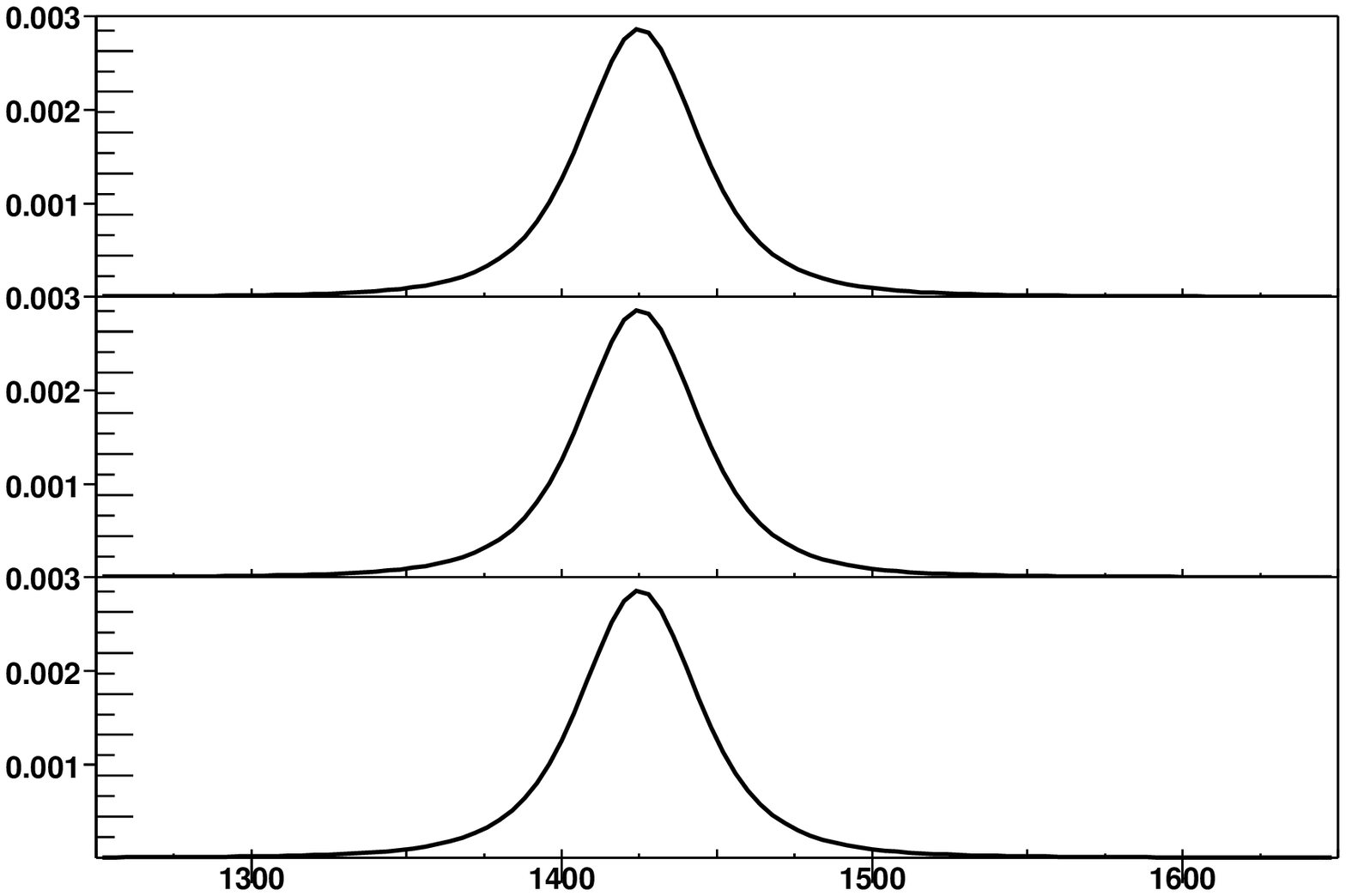}&
\hspace*{-8mm}\includegraphics[width=0.36\textwidth,height=6cm]{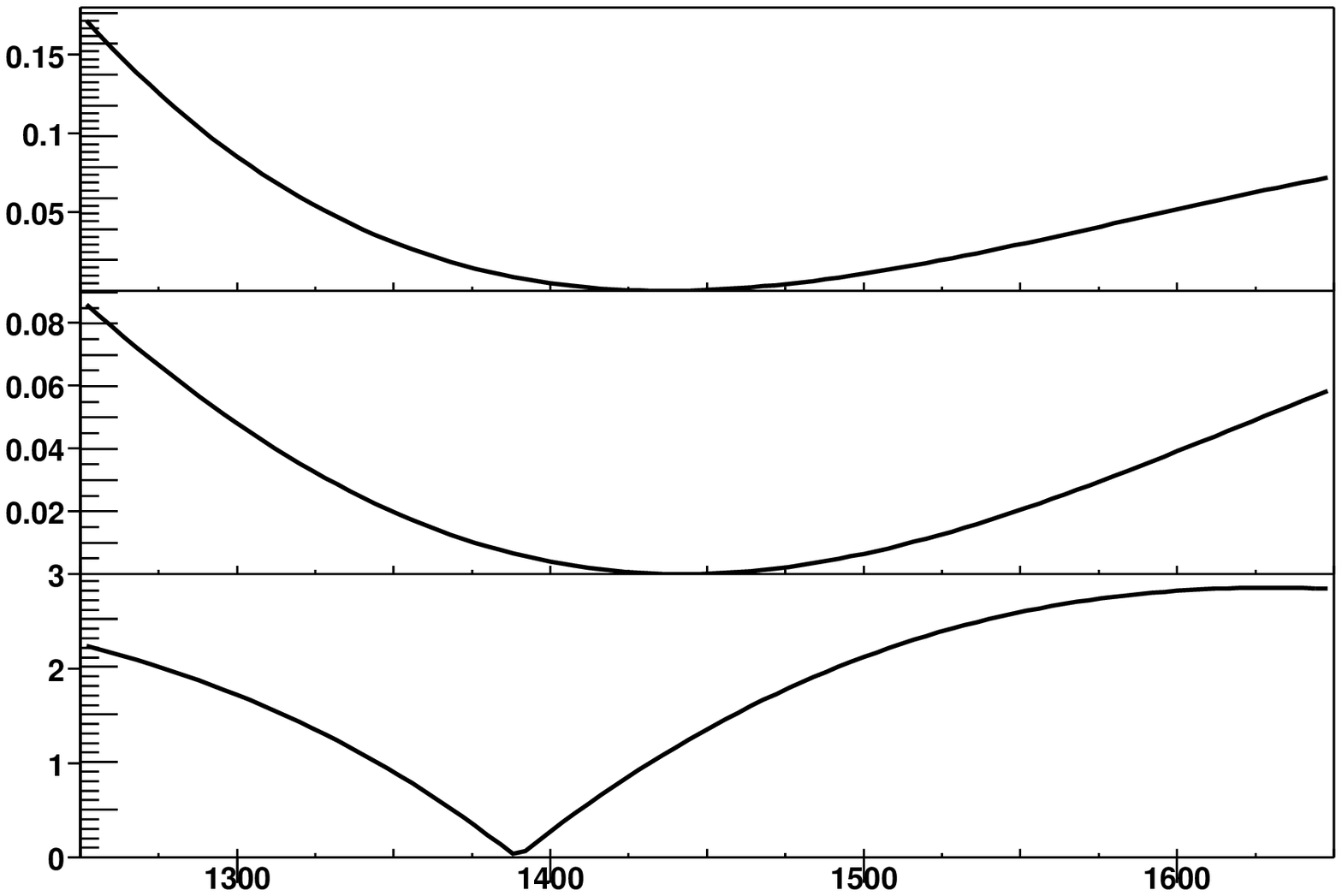}&
\hspace*{-8.5mm}\includegraphics[width=0.38\textwidth,height=6cm]{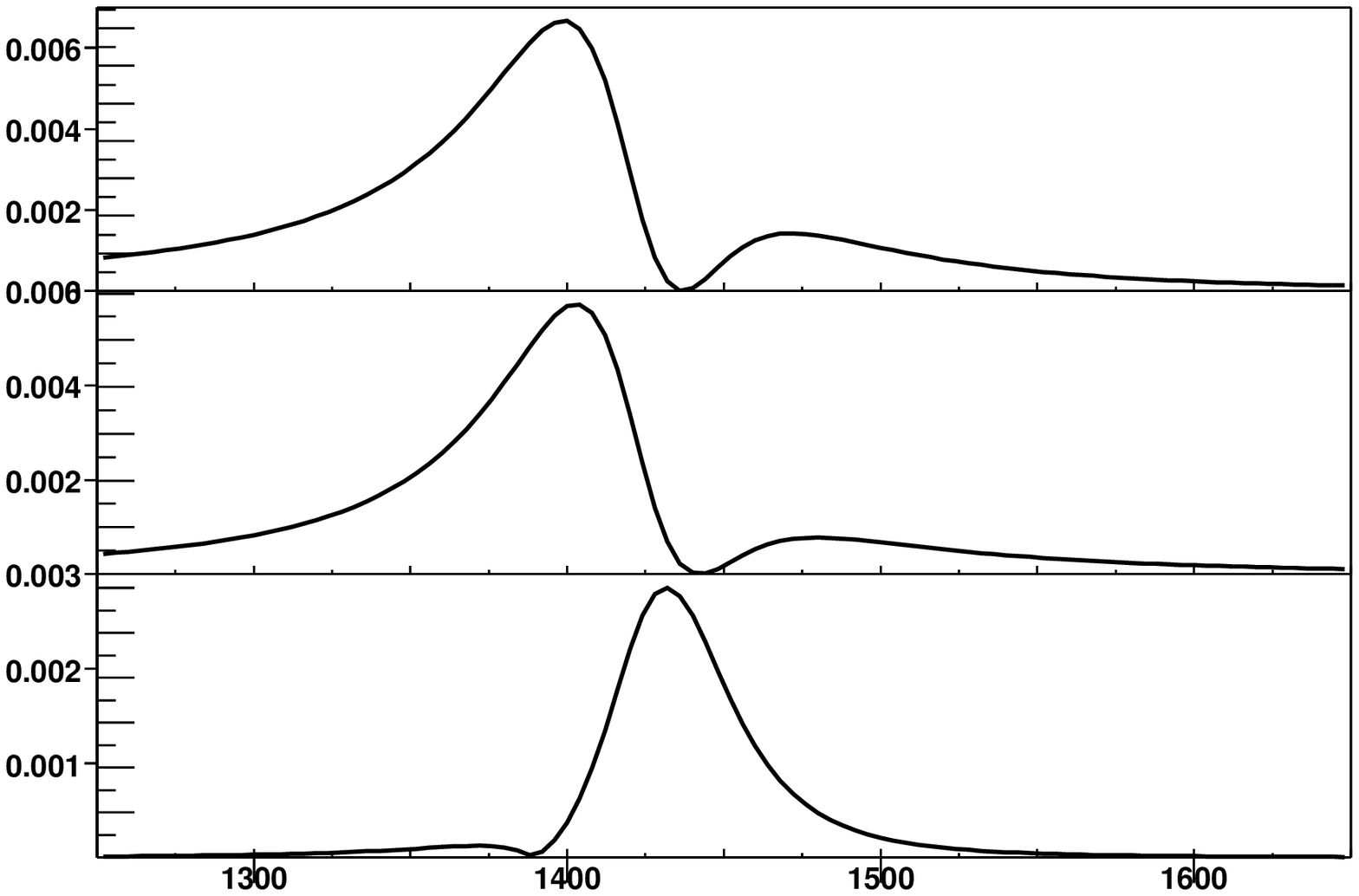}
\end{tabular}
\caption{\label{node}
Amplitudes for $\eta(1440)$ decays to $a_0\pi$ (first row), 
$\sigma\eta$ (second row), and $\rm K^*\bar K$ (third row);
the Breit-Wigner functions are shown on the left, then the squared 
decay amplitudes~\protect\cite{Barnes:1996ff} and, on the right, 
the resulting squared transition matrix element.}
\end{figure}

The matrix element for $\eta_R$ decays to $a_0(980)\pi$ or
$\sigma\eta$ has the form
$$
f_S = \frac{2^{4}}{3^{4}}\cdot \left(1-\frac{7}{9}x^2 +
\frac{2}{27}x^2 \right)
$$
and vanishes for  $p=0.45$\,GeV/c.  
The decay to $a_0(980)\pi$
vanishes at the mass 1440\,MeV. This has a decisive impact
on the shape, as seen in Figure~\ref{node}. Shown are 
the transition matrix elements as given by Barnes et al.~\cite{Barnes:1996ff}
and the product of the squared matrix elements and a Breit--Wigner
distribution with mass 1420\,MeV and  width 60\,MeV.
\par
\par
The $\eta(1440)\to a_0(980)\pi$ and $\to\rm K^*K$ mass distributions
have different peak positions;  at approximately
the $\eta_L$ and $\eta_H$ masses. Hence there is no need to introduce
the $\eta_L$ and $\eta_H$ as two independent states. One
$\eta(1420)$ and the assumption that it is a radial excitation
describe the data. 
\par
\begin{figure}[htb]
\begin{tabular}{cc} 
\includegraphics[width=0.5\textwidth]{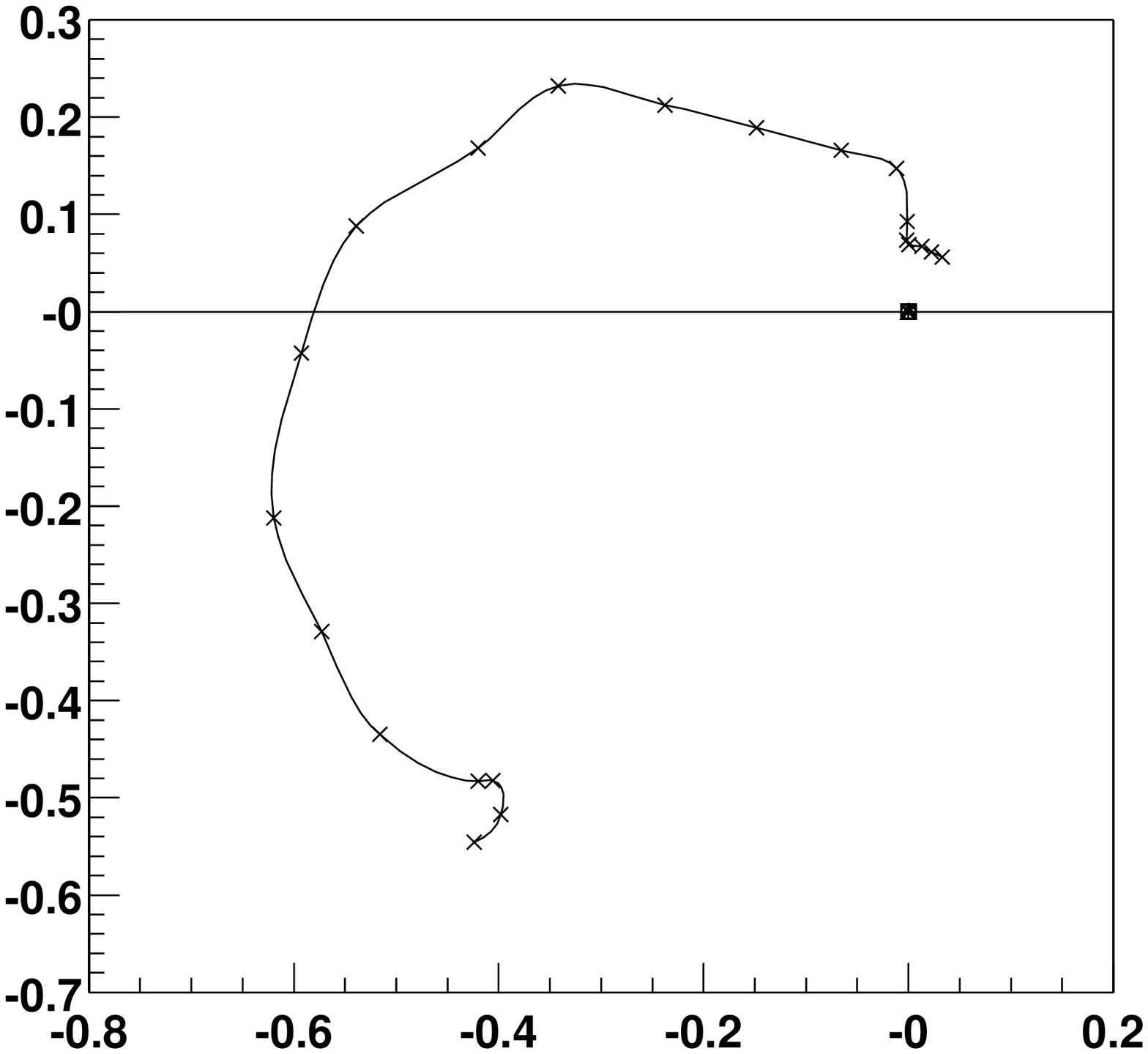}&\hspace*{-8mm}
\includegraphics[width=0.5\textwidth]{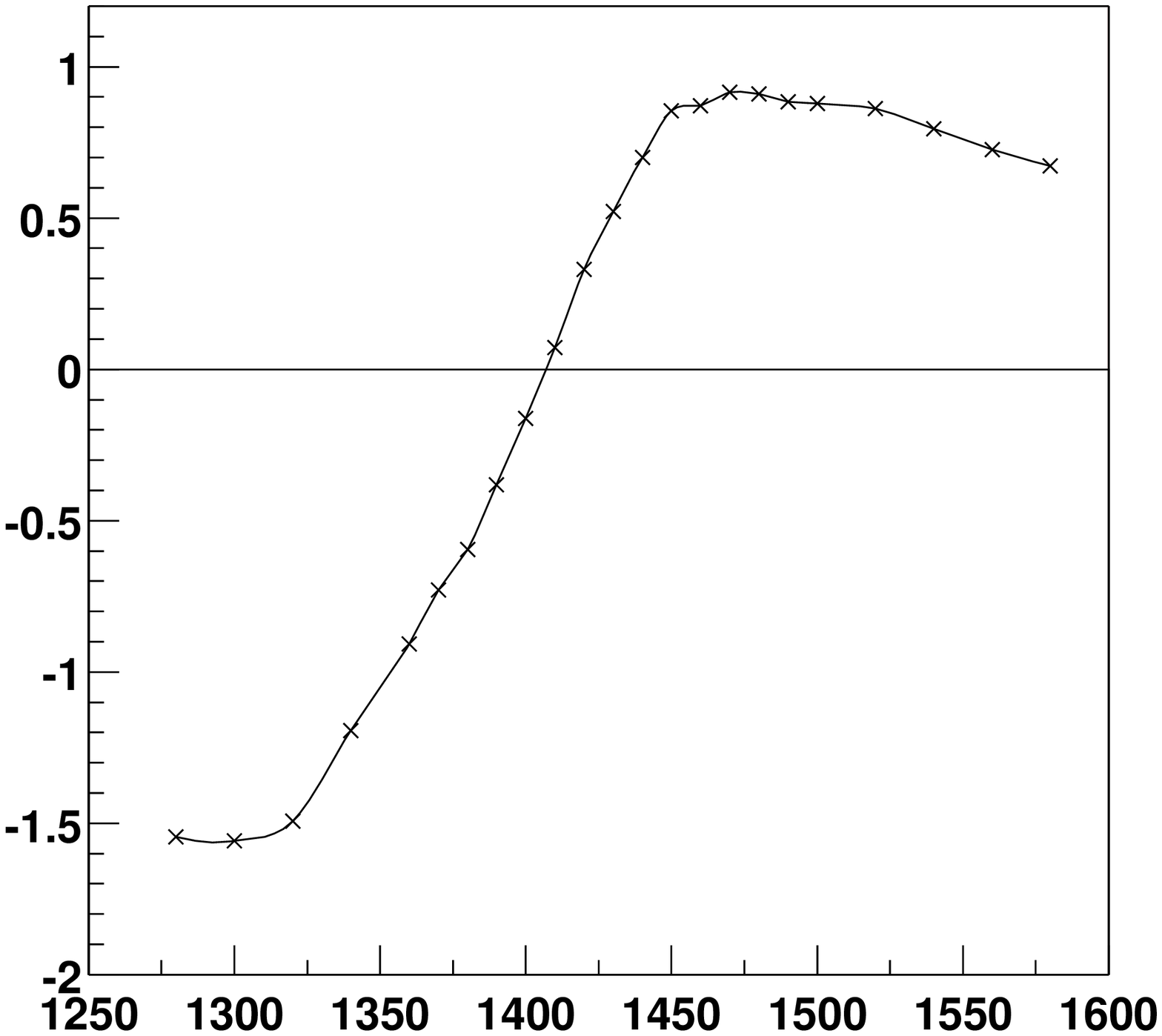}
\end{tabular}
\caption{\label{phase}
Complex amplitude and phase motion of the 
$a_0(980)\pi$ isobars
in $\rm p\bar p$ annihilation into $4\pi\eta$. In the mass range
from 1300 to 1500\,MeV the phase
varies by $\pi$ indicating that there is only one resonance in the
mass interval. The $\sigma\eta$ (not shown) exhibits 
the same behavior~\protect\cite{Reinnarth}.
}
\end{figure}

This conjecture can be
further tested by following the phase motion of the
$a_0(980)\pi$ or $\sigma\eta$ isobar~\cite{Reinnarth}. 
The phase changes by $\pi$ and
not by 2$\pi$, see Fig.~\ref{phase}.
\par
\par
\section{Conclusions}
Summarizing, the results for the radial excitations of
pseudoscalar mesons are as follows:
\begin{itemize}
\item The $\eta(1295)$ is not a $q\bar q$ meson.
\item The $\eta (1440)$ wave function
has a node leading to two appearantly different states $\eta_L$
and $\eta_H$. 
\item There is only one $\eta$ state, the $\eta(1420)$,
in the mass range from 1200 to 1500 MeV and not 3\,!
\item
The $\eta(1440)$ is the radial excitation
of the $\eta$. 
The radial excitation of the $\eta'$ is expected
at about 1800\,MeV; it might be the $\eta (1760)$.
\end{itemize}

The following states are most likely the pseudoscalar
ground states and radial excitations:

\begin{center}
\renewcommand{\arraystretch}{1.3}
\begin{tabular}{lcccc}
\hline\hline
$1^1S_0$& $\pi$ & $\eta^{\prime}$ & $\eta$ & K  \\
$2^1S_0$\hspace*{-2mm}&\hspace*{-2mm} $\pi(1300)$ \hspace*{-2mm}&\hspace*{-2mm} $\eta(1760)$ \hspace*{-2mm}&\hspace*{-2mm} $\eta(1440)$ &\hspace*{-2mm} K(1460)\hspace*{-2mm} \\
\hline\hline
\end{tabular}
\renewcommand{\arraystretch}{1.3}
\end{center}

\end{document}